\documentclass[12pt,english,journal,draftclsnofoot,onecolumn]{IEEEtran}

\usepackage[dvips]{graphicx}
\usepackage{psfrag}
\usepackage[T1]{fontenc}
\usepackage[latin9]{inputenc}
\usepackage{float}
\usepackage{amsthm}
\usepackage{amsmath}
\usepackage{amssymb}
\usepackage{graphicx}
\usepackage{esint}

\makeatletter
\theoremstyle{plain}

\usepackage{amsthm}
\theoremstyle{plain}
\@ifundefined{definecolor}{\usepackage{color}}{}
\usepackage{algorithmic}\usepackage{algorithm}

\IEEEoverridecommandlockouts

\makeatother

\usepackage{babel}
\providecommand{\theoremname}{Theorem}
\addto\captionsenglish{}

\begin{document}

\title{Low Power Wideband Sensing for One-Bit Quantized Cognitive Radio Systems}

\author{{\normalsize {Abdelmohsen Ali and Walaa Hamouda}} \thanks{%
A. Ali and W. Hamouda with the Dept. of Electrical and Computer Engineering, Concordia University, Montreal, Quebec, H3G 1M8, Canada e-mail:\{(ali\_abde,hamouda)@ece.concordia.ca\}}}
\maketitle

\begin{abstract}
We proposes an ultra low power wideband spectrum sensing architecture by utilizing a one-bit quantization at the cognitive radio (CR) receiver. The impact of this aggressive quantization is quantified and it is shown that the proposed method is robust to low signal-to-noise ratios (SNR). We derive closed-form expressions for both false alarm and detection probabilities. The sensing performance and the analytical results are assessed through comparisons with respective results from computer simulations. The results indicate that the proposed method provides significant saving in power, complexity, and sensing period on the account of an acceptable range of performance degradation.

\end{abstract}


\IEEEpeerreviewmaketitle


\section{Introduction}


Wideband spectrum sensing, which consists of observing a wideband and identifying the portions that are occupied and those which are free, is essential in interweave cognitive radio networks [1]. In these networks, unlicensed or secondary users (SUs) are prohibited from accessing an occupied band by the primary user (PU). The SU has to vacate the band and search for another unoccupied band if the PU appears, thus wideband sensing is a must. One of the main approaches to realize wideband sensing is to assume the feasibility of sampling the desired spectrum by the ordinary Nyquist rate [2]. Practically, high computational complexity and power consumption attached to the required ultra high sampling rates and high Analogue to Digital Converter (ADC) resolutions can be relaxed if the sensing performance is acceptable with ultra low precision ADCs (1-3 bits) [3].


In this work, we propose a wideband spectrum sensing system in which a 1-bit ADC is employed. In fact, among all other ADCs, a 1-bit ADC consumes the minimum power for a given sampling frequency. Current commercially available high-resolution high-speed ADCs consume power on the order of several Watts. For example, the recent 12-bit ADC 12D1600QML-SP from Texas Instruments can process 3.2GSamples/sec at a power consumption of 3.88Watts. On the other hand, a single high speed comparator (1-bit ADC) with the same operating frequency is designed to dissipate 20$\mu$Watts [4]. Further, the complexity is extremely reduced as automatic gain control is not required for these systems. Also the hardware complexity for the digital signal processing modules including the Fast Fourier Transform (FFT) engine and the power detector will be significantly reduced due to the minimized bit-width for various building blocks. For instance, the FFT module will not involve any multiplications and only additions/subtractions are required. Motivated by the above, we present a complete architecture for the sensing engine. Irrespective of the sensing period being long, the system performance for both non-quantized and 1-bit quantized systems is modelled analytically and evaluated by simulations.


\section{System Model and Assumptions}
\label{sec:systemmodel}

Consider a CR system operating over a wideband channel divided into $N$ non-overlapping sub-bands. Conventionally, it is assumed that the sub-bands have equal-size bandwidths $B$ [5]. At the CR receiver, the signal is sampled at the Nyquist sampling rate $F_s=NB$. The received signal in the frequency domain can be represented by the $1 \mathrm{\times} N$ vector given by,
\begin{eqnarray}
\mathbf{R}&=& \sum_{m=1}^{M} \mathbf{H}_m \mathbf{S}_m+\mathbf{W} \label{equ:RxDiscreteFDsignal}
\end{eqnarray}
\noindent where $M$ represents the number of active primary signals that randomly occupy $M$ sub-bands and $M<N$, $\mathbf{H}_m$ is a diagonal $N \times N$ channel matrix, $\mathbf{S}_m$ is the spectrum of the $m$th primary signal over the $m$th sub-band, and $\mathbf{W}$ is frequency domain independent and identically distributed circularly-symmetric additive white Gaussian noise with zero-mean and variance $\mathbb{E}[\mathbf{W}\mathbf{W}^T]=N\sigma^2_W$, where $\mathbb{E}$ denotes expectation. Similar to [6], it is assumed that the distribution of the received primary signal over a single sub-band is also circularly-symmetric complex Gaussian (CSCG) with zero-mean and variance $\sigma^2_S$. This assumption holds when primary radios deploy uniform power transmission strategies given no channel knowledge at the transmitter side [2]. That is the total power over the entire band is $\mathbb{E}[\mathbf{R}\mathbf{R}^T]\,=\,M\sigma^2_S+N\sigma^2_W$ where all interferers and the noise are assumed to be statistically independent. Finally, the channel is assumed to be static over the sensing interval.


\section{Proposed Spectrum Sensing Procedure}
\label{sec:ProposedProcedure}

\begin{figure*}[!t]
\centering\includegraphics[scale=0.47]{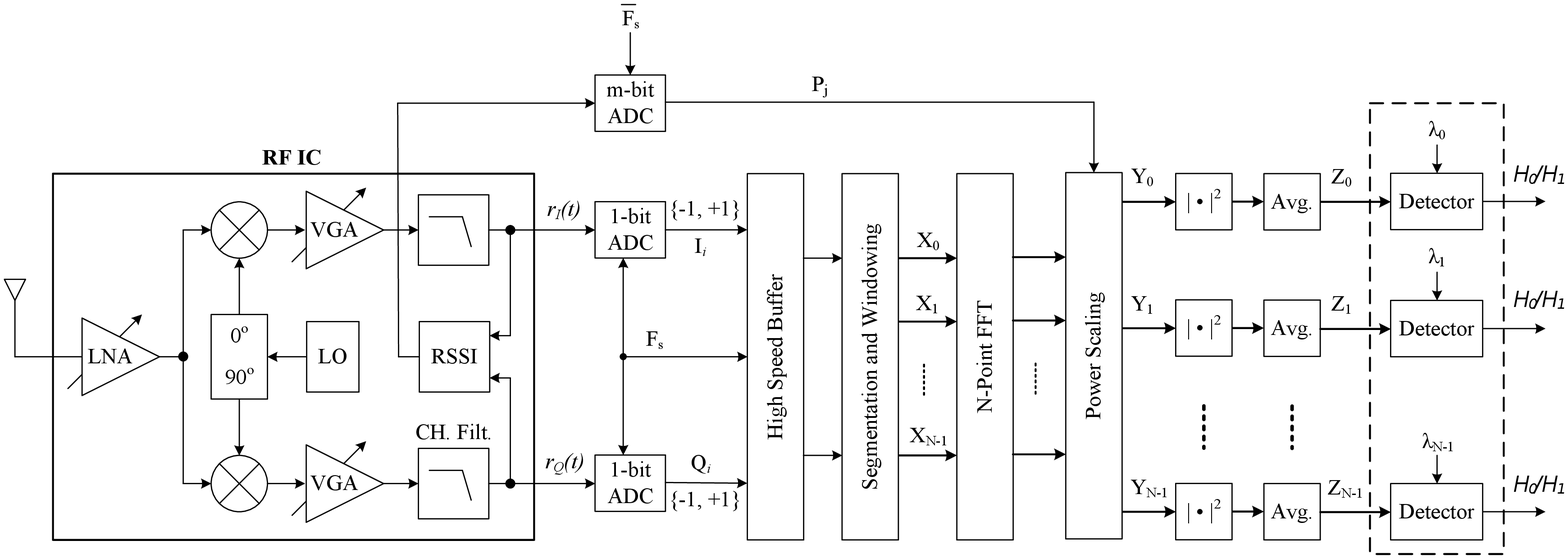}
\caption{General architecture for the low power wideband sensing system involving 1-bit quantizers.}
\label{fig:OneBitQuantizerRSSI}
\end{figure*}

The architecture for the proposed one-bit quantized system is shown by Fig.~\ref{fig:OneBitQuantizerRSSI}. After the RF direct conversion processing, the baseband signal is sampled through a 1-bit quantizer to retrieve the sign from the sample value. A high speed buffer is employed to store one window of captured samples, precisely $LN$ samples. The samples are segmented into non-overlapped captures where each capture has exactly $N$ samples. Current RF processors fortunately provide the received signal power that is measured by a received signal strength indicator (RSSI) block [7]. The measured value is reported to the digital processor through a stand-alone high resolution ADC that operates at relatively low sampling rate, $\bar{F}_s$. In this work, each capture is processed by the FFT block to obtain the frequency spectrum over $N$ frequency pins where each pin corresponds to a single sub-band. The frequency transformation and scaling can be defined by,
\begin{eqnarray}
Y_{n,i}&=&\sqrt{\frac{P_j}{2N}} \; \sum_{k=0}^{N-1} X_{k,i}\,e^{-\mathbf{j}2\pi kn/N},\;\mathbf{j}=\sqrt{-1} \label{equ:FDsignal}
\end{eqnarray}

\noindent where $i$ is the capture index, $n$ is the sub-band index such that $0\leq n<N$, $P_j$ is the measured signal power, and $j$ is an integer representing the window index over time. According to our model, the measured power value corresponds to $P_j=M\sigma_S^2+N\sigma_W^2$. For each frequency pin $n$ and window index $j$, the energy contained in one window consisting of $L$ samples can be defined by $T_{n,j}$ as given by (\ref{equ:DecisionStatistic}). Hence, the decision statistic, $Z_{n,j}=T_{n,j}/L$, is defined to be a simple power estimator that is employed to decide whether this sub-band is a hole or not.
\begin{eqnarray}
T_{n,j}&=&\sum_{k=0}^{L-1} \Big|Y_{n,j\times L+k}\Big|^2 \label{equ:DecisionStatistic}
\end{eqnarray}

\section{Energy Detection Performance For One Bit Quantizer System}
\label{sec:ERanalysis1Bit}

For a large number of samples to average, the central-limit theorem is typically employed to approximate the probability distribution function (PDF) of the decision statistic $Z_{n,j}$ as a normal distribution under both hypothesises [6]. In this section, we provide exact closed-form expressions for the sensing performance independently on the averaging depth by considering first the non-quantized system. Next, the effect of the 1-bit ADC is demonstrated by evaluating the amount of noise power added to the frequency spectrum for both hypothesises.

\subsection{Non-quantized Exact System Performance}

For each sub-band, $n$, we wish to discriminate between the two hypotheses $\mathcal{H}_{0,n}$ and $\mathcal{H}_{1,n}$ where the first assumes that the primary signal is not in band and the second assumes that the primary user is present. Using the average energy decision statistic, one can define these hypotheses under the assumption of infinite ADC precision as given by (\ref{equ:BinaryHypotheses}).
\begin{eqnarray}
&& \left\{
  \begin{array}{l l}
    \mathcal{H}_{0, n}:\;Z_{n,j} \leq \lambda_n,\, & \sigma_0^2\,=\,\sigma_Y^2\,=\,\sigma_W^2 \\
    \mathcal{H}_{1, n}:\;Z_{n,j}>\lambda_n,\, & \sigma_1^2\,=\,\sigma_Y^2\,=\,\sigma_S^2+\sigma_W^2\\
  \end{array} \right.
  \label{equ:BinaryHypotheses}
\end{eqnarray}

As the wideband sensing objective is to explore the spectral occupancy of primary signals over numerous number of sub-bands (e.g., $N\gg 100$), the FFT output sequence follows a CSCG distribution by the central-limit theorem. Let $Y_{n,i} \sim \mathcal{CN}(0, \sigma_Y ^2)$, the random variable $T_{n,j}$ follows a Chi-square distribution with $L$ degrees of freedom [8]. By applying a linear transformation between random variables, one can obtain the PDF for the decision statistic as given by (\ref{equ:EnergydistributionforZ}) where $\sigma_Z ^2=\sigma_Y ^2/L$. Further, the cumulative distribution function (CDF) can be obtained in a closed-form as given by (\ref{equ:EnergydistributionCDF}).
\begin{eqnarray}
f_{Z_{n,j}}(z)&=& \frac{1}{\sigma_Z ^{2L}\,\Gamma(L)}\,z^{L-1}\,e^{-z/\sigma_Z ^2}\, ,\;\; z>0 \label{equ:EnergydistributionforZ} \\
F_{Z_{n,j}}(z)&=& 1- \sum_{k=0}^{L-1} {\frac{1}{k!}\,\bigg(\frac{z}{\sigma_Z ^2}\bigg)^{k}}\,e^{-z/\sigma_Z ^2} \label{equ:EnergydistributionCDF}
\end{eqnarray}

The quality of the detector is described by the Receiver-Operating-Characteristics (ROC) which represent the probability of detection, $P_D$, and the probability of false alarm, $P_{FA}$, that are defined as the probabilities that the sensing algorithm detects a primary user under hypotheses $\mathcal{H}_{1, n}$ and $\mathcal{H}_{0, n}$, respectively. By varying a certain threshold $\lambda_n$ for each sub-band $n$, the operating point of a detector can be chosen anywhere along the ROC curve. $P_{FA}$ and $P_D$ can be defined as given by (\ref{equ:PFAdefinition}) and (\ref{equ:PDdefinition}), respectively.
\begin{eqnarray}
\setcounter{equation}{9}
\hrulefill
P_{FA}&=& Prob\big[Z_{n,j}> \lambda_n \,|\,\mathcal{H}_0 \big] \,=\,1-F_{Z_{n,j}|\mathcal{H}_{0,n}}(\lambda_n) \nonumber \\
&=& \sum_{k=0}^{L-1} {\frac{1}{k!}\,\bigg(\frac{\lambda _n L}{\sigma_W ^2}\bigg)^{k}}\,e^{-\lambda _n L/\sigma_W ^2} \label{equ:PFAdefinition} \\
P_{D}&=& Prob\big[Z_{n,j}> \lambda_n \,|\,\mathcal{H}_1 \big] \,=\,1-F_{Z_{n,j}|\mathcal{H}_{1,n}}(\lambda_n) \nonumber \\
&=& \sum_{k=0}^{L-1} {\frac{1}{k!}\,\bigg(\frac{\lambda _n L}{\sigma_W ^2+\sigma_S ^2}\bigg)^{k}}\,e^{-\lambda _n L/(\sigma_W ^2+\sigma_S ^2)}\label{equ:PDdefinition}
\end{eqnarray}

\subsection{One-Bit Quantization Performance}

In conventional systems that consider the quantization effect [9], the effect is modelled by adding one more term to the signal variance representing the quantization noise power which is a function of the ADC resolution. Unfortunately, this procedure cannot be applied for the 1-bit quantizer case since the ADC aggressively saturates the incoming signal to two possible outcomes $\{-1, +1\}$ that are uniformly distributed.

By introducing the power scaling operation after the FFT module, a total power transfer is guaranteed to the frequency domain since the defined transform itself is linear and unitary. However, the main objective of the transformation is to reshape the power across various sub-bands. If the input is left un-quantized, the information required for this redistribution process is known in full and the detection error is only introduced due to the noisy environment. When the input is quantized to a single bit and no power gain or loss is guaranteed, then simply the quantization effect can be interpreted as a power leakage process due to the reduced amount of information about the power distribution. Let $\alpha M\sigma_S^2$ be the amount of leakage to all sub-bands where $\alpha$ is a constant. Since the occupied sub-bands are uniformly distributed across the whole band, the leakage will also be uniformly distributed across all sub-bands. Then, the power contained by one sub-band under $\mathcal{H}_{0, n}$ would be $\sigma_0 ^2=\sigma_W ^2(1+\alpha \gamma M/N)$ where $\gamma=\sigma_S^2/\sigma_W^2$ is the SNR over one sub-band. One part of the leaked power is distributed over the vacant sub-bands while the remaining part is added to the occupied sub-bands themselves. Thus, one can write the power contained in one occupied sub-band as $\sigma_1 ^2=\sigma_W ^2(1+\gamma-\gamma \alpha+\alpha \gamma M/N)$ where $-\gamma \alpha$ represents the contribution of this sub-band in the total leakage power. In this work, we rely on extensive computer simulations by varying $\sigma_W^2$, $\sigma_S^2$, $M$, $L$, and $N$ to find an optimum estimate for this constant which is found to be $\alpha \simeq e^{-1}$.

\section{Simulation Results}
\label{sec:Results}

\begin{figure}[t]
\centering\includegraphics[scale=0.59]{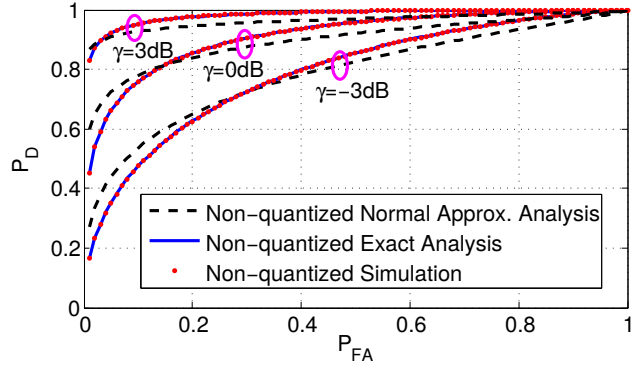}
\caption{ROC comparison among the non-quantized exact analysis, the corresponding approximated analysis~\cite{ref:equalTxPower}, and the non-quantized simulations at different SNR values for $L=8$ and $M=100$}
\label{fig:ROCunquanNormalquanCompare2Mod}
\end{figure}
In the simulation, we consider a wideband system that employs a total band of $1024$MHz which is divided into $N=1024$ non-overlapped sub-bands, $M$ of which are occupied by primary signals. Each of those allocated sub-bands carries QAM signal that is passed over a blocking fading channel filter. To simulate the system behaviour, $10^5$ trials are processed and the system performance is evaluated based on the decision outcomes. In each trial, a single window is generated where the sub-band occupancies are never changed within a single window. The signs of the received samples are captured to be processed by the detector.

First, the approximated ROC proposed by [6] is compared to our exact closed-form ROC performance under different SNR values and for relatively high averaging rate (e.g, $L=8$). The performance curves are shown in Fig.~\ref{fig:ROCunquanNormalquanCompare2Mod}. It is clear that the Normal approximation introduces considerably large errors in performance even for relatively high averaging rate. The simulation results for the non-quantized system are also shown to demonstrate the accuracy of our derivations.

\begin{figure}[t]
\centering\includegraphics[scale=0.54]{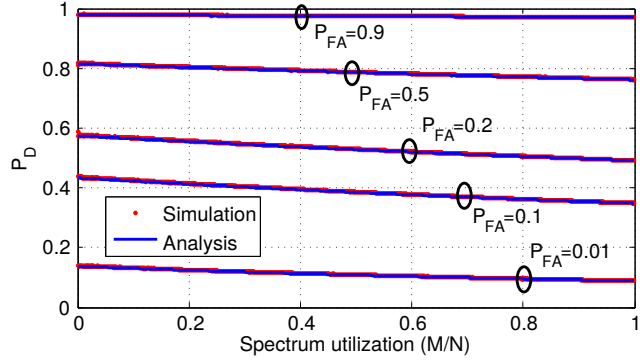}
\caption{Detection performance for the one bit quantizer versus the spectrum utilization at $\gamma=0$dB, $L=4$, and different $P_{FA}$ values.}
\label{fig:DetectionDiffMSNR0Quan2Mod}
\end{figure}

\begin{figure}[t]
\centering\includegraphics[scale=0.55]{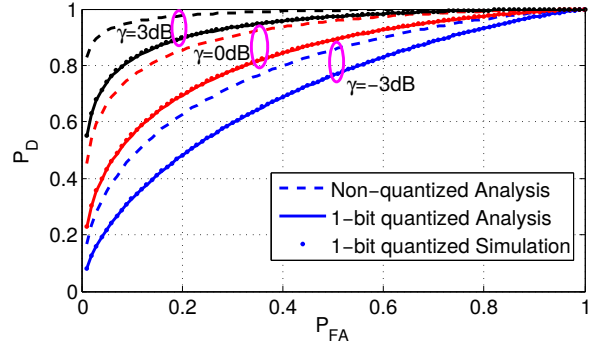}
\caption{ROC comparison for different SNR values ($\gamma=$-3, 0, or 3 dB) for $L=8$ and $M=100$}
\label{fig:ROCAnalysisSimQuanUnquanCompare2Mod}
\end{figure}

\begin{figure}[t]
\centering\includegraphics[scale=0.45]{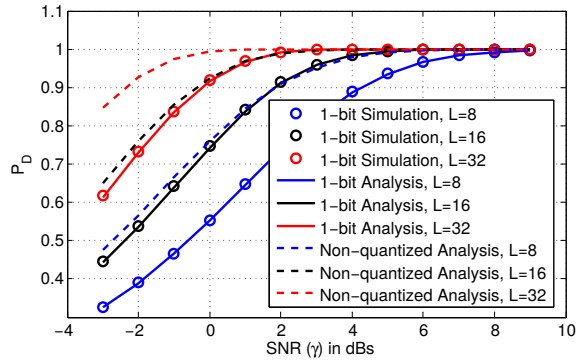}
\caption{Detection probability for the non-quantized exact analysis, the 1-bit quantized analysis, and the 1-bit quantized simulation for different time average values for $P_{FA}=0.1$ and $M=100$}
\label{fig:PDUnquanQuanDiffLCompare2Mod}
\end{figure}

Next, extensive simulations are performed to verify the constant value $\alpha \simeq e^{-1}$. In these results, more than 100 false alarm rates are simulated and different SNR values are considered. Fig.~\ref{fig:DetectionDiffMSNR0Quan2Mod} shows the exact match of the performance between the simulation and the analysis for all possible spectrum utilization ratios and for countable number of false alarm rates. Although these results assume fixed values for other parameters such as $L=4$ and SNR=0dB, we rely on other results (in Fig.~\ref{fig:ROCAnalysisSimQuanUnquanCompare2Mod} and Fig.~\ref{fig:PDUnquanQuanDiffLCompare2Mod}) to demonstrate the confidence and the effectiveness of our selected constant value.

Fig.~\ref{fig:ROCAnalysisSimQuanUnquanCompare2Mod} shows the ROC for the 1-bit quantizer and the non-quantized systems. We emphasize the fact that the simulations exactly match the analysis for the 1-bit quantizer case for various SNR cases and for a different averaging rate $L=8$ than the one used in Fig.~\ref{fig:DetectionDiffMSNR0Quan2Mod}. The quantization effect is clear in Fig.~\ref{fig:PDUnquanQuanDiffLCompare2Mod} where a degradation of about 2dB is observed. By allowing more averaging time, the degradation can be improved.


\section{Conclusion}
\label{sec:conclusion}
We proposed a 1-bit quantization architecture for wideband spectrum sensing in interweave cognitive radio networks. The ultimate goal is to extremely reduce the power consumption and complexity while the activity of primary users is detected with relatively high accuracy. We derived the exact non-quantized ROC independent of the sensing interval. Further, we provided analytical expressions for the false alarm and detection rates for the proposed 1-bit quantizer. Simulation results indicate that the derivations are accurate and reliable for various system parameters.

\section*{References}
\begin{small}

\noindent [1] L. De Vito, "A review of wideband spectrum sensing methods for Cognitive Radios," in \emph{IEEE Inter. Instrumentation and
Measurement Tech. Conf.}, May 2012, pp. 2257-2262.

\noindent [2] Z. Quan et al., "Optimal Multiband Joint Detection for Spectrum Sensing in Cognitive Radio Networks," \emph{IEEE Trans. on
Signal Processing}, vol. 57, no. 3, pp. 1128-1140, March 2009.

\noindent [3] J. Laska, Z. Wen, W. Yin, and R. Baraniuk, "Trust, But Verify: Fast and Accurate Signal Recovery From 1-Bit Compressive
Measurements," \emph{IEEE Transactions on Signal Processing}, vol. 59, no. 11, pp. 5289-5301, Nov 2011.

\noindent [4] V. Bhumireddy et al., "Design of low power and high speed comparator with sub-32-nm Double Gate-MOSFET," in \emph{Inter.
Conf. on Circuits and Systems}, Sept. 2013, pp. 1-4.

\noindent [5] T.-H. Yu, S. Rodriguez-Parera, D. Markovic, and D. Cabric, "Cognitive Radio Wideband Spectrum Sensing Using Multitap
Windowing and Power Detection with Threshold Adaptation," in \emph{IEEE Inter. Conf. on Comm.}, May 2010, pp. 1-6.

\noindent [6] Y.-C. Liang, Y. Zeng, E. Peh, and A. T. Hoang, "Sensing-Throughput Tradeoff for Cognitive Radio Networks," \emph{IEEE
Transactions on Wireless Communications}, vol. 7, no. 4, pp. 1326-1337, April 2008.

\noindent [7] N. Bambal and S. Dixit, "CMOS Limiting Amplifier and RSSI (Received Signal Strength Indicator)," in \emph{Inter. Conf. on
Emerging Trends in Eng. and Tech.}, Nov 2011, pp. 238-243.

\noindent [8] F. Digham, M.-S. Alouini, and M. K. Simon, "On the energy detection of unknown signals over fading channels," in \emph{IEEE
Inter. Conf. on Comm.}, vol. 5, 2003, pp. 3575-3579.

\noindent [9] Z. Liu, Y. Liu, and H. Yang, "Energy efficient A/D conversion for sequential wideband multichannel spectrum sensing in
cognitive radio network," in \emph{International Conference on Advanced Communication Technology}, Jan 2013, pp. 580-585.

\end{small}
\end{document}